\documentclass{INTERSPEECH2023}

\usepackage{dcolumn}
\usepackage{multirow}
\usepackage{hyperref}
\usepackage[table]{xcolor}
\usepackage{fontawesome5}

\newcommand*\rot{\rotatebox[origin=c]{90}}

\newcommand*\variant{\hspace{0.1cm}\rotatebox[origin=c]{180}{$\Lsh$}\hspace{0.1cm}}
\newcommand*\subvariant{\hspace{0.5cm}\rotatebox[origin=c]{180}{$\Lsh$}\hspace{0.1cm}}

\DeclareMathOperator*{\argmax}{arg\,max}

\newcommand{\smalltexttt}[1]{{\scriptsize\texttt{#1}}}

\interspeechcameraready 

\title{Powerset multi-class cross entropy loss for neural speaker diarization}
\name{Alexis Plaquet \& Hervé Bredin}
\address{IRIT, Université de Toulouse, CNRS, Toulouse INP, UT3, Toulouse, France}
\email{firstname.lastname@irit.fr}

\begin{document}

\maketitle
 
\begin{abstract}
Since its introduction in 2019, the whole end-to-end neural diarization (EEND) line of work has been addressing speaker diarization as a frame-wise multi-label classification problem with permutation-invariant training. Despite EEND showing great promise, a few recent works took a step back and studied the possible combination of (local) supervised EEND diarization with (global) unsupervised clustering. Yet, these hybrid contributions did not question the original multi-label formulation. We propose to switch from multi-label (where any two speakers can be active at the same time) to powerset multi-class classification (where dedicated classes are assigned to pairs of overlapping speakers). Through extensive experiments on 9 different benchmarks, we show that this formulation leads to significantly better performance (mostly on overlapping speech) and robustness to domain mismatch, while eliminating the detection threshold hyperparameter, critical for the multi-label formulation.
\end{abstract}

\section{Introduction}

Speaker diarization is the task of partitioning an audio stream into homogeneous temporal segments according to the identity of the speaker. Most dependable diarization approaches consist of a cascade of several steps~\cite{vbx}: voice activity detection to discard non-speech regions, speaker embedding to obtain discriminative speaker representations, and clustering to group speech segments by speaker identity. This family of approaches has two main drawbacks: 
\begin{itemize}
    \item errors made by each step are propagated to the subsequent steps, possibly escalating into even larger errors;
    \item they are not designed to detect (let alone assign to the right speakers) overlapping speech regions -- an additional post-processing step is needed to handle them~\cite{Bullock2020}.
\end{itemize}

To circumvent these limitations, a new family of approaches have recently emerged, rethinking speaker diarization completely~\cite{Fujita2019,eend_attention,Bredin2021}. Dubbed end-to-end diarization (EEND), the main idea is to train a single neural network that ingests the audio recording and directly outputs the diarization, hence addressing the error propagation issue. They formulate the problem as a \textit{multi-label classification} task, allowing multiple overlapping speakers to be active simultaneously. Furthermore, \textit{permutation-invariant training}~\cite{Fujita2019} is the critical ingredient that turns clustering -- an intrinsically unsupervised task -- into a supervised classification task. EEND approaches do have a few significant limitations:
\begin{itemize}
    \item they struggle to predict the correct number of speakers, especially when it is larger in test data than during training -- additional mechanisms such as encoder-decoder attractors were proposed to partially solve this issue~\cite{Horiguchi2020}; 
    \item training such approaches is extremely data hungry as each conversation in the training set constitutes a single training sample -- one has to resort to synthetic (hence unrealistic) conversations instead~\cite{Landini2022};
    \item because of the internal self-attention mechanism, they do not scale well to long conversations.
\end{itemize}

Therefore, a few recent works took a step back to borrow the \textit{best of both worlds} (BoBW, \textit{i.e.} multi-stage and end-to-end approaches)~\cite{EEND-clustering, EEND-clustering-advances}. The general principle is based on three steps:
\begin{enumerate}
\item split long conversations into shorter chunks;
\item apply end-to-end speaker diarization on each of them;
\item stitch them back together using speaker embeddings and unsupervised clustering.
\end{enumerate}

To both start with a strong baseline and ensure reproducibility, we build on top of the speaker diarization pipeline available in version {\scriptsize \texttt{2.1.1}} of {\scriptsize \texttt{pyannote.audio}} open source toolkit \cite{pyannote2} 
that embraces this aforementioned three-steps principle. More precisely, we focus on the loss function used for training neural networks towards end-to-end speaker diarization of short audio chunks. \cite{Bredin2021} calls this task \textit{speaker segmentation} and states that working with short audio chunks has the following advantages:
\begin{itemize}
    \item it circumvents the scalability issue because the neural network only ever ingests short fixed-duration (5s in our case) audio chunks;
    \item it makes it easier to train (as each conversation can now provide multiple training samples);
    \item it allows to set a small upper bound on the (local) number of speakers $K_\text{max}$ -- for instance, there is a 99\% chance that any 5s chunk contains less than $K_\text{max}=3$ speakers in the 9 benchmarking datasets used in this paper.
\end{itemize}

\begin{figure}[t]
  \centering
  \includegraphics[width=0.8\linewidth]{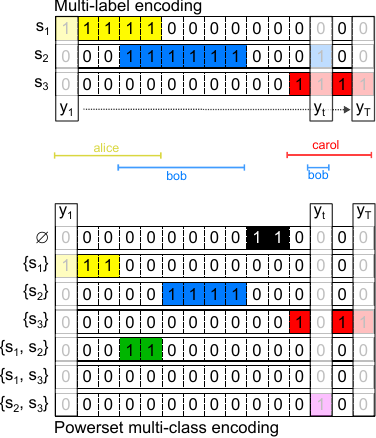}
  \caption{From multi-label to powerset multi-class encoding. While \textit{powerset} classes are mutually exclusive, several multi-label classes must be active at the same time to encode overlapping speakers.}
  \label{fig:encoding}
\end{figure}

Since its introduction a few years ago, the EEND line of work has been processing whole conversations with up to $K_\text{max} \approx 20$ (partially overlapping) speakers. Therefore, addressing it as a frame-wise multi-label classification problem was a brilliant (and efficient!) idea that we summarize in Section~\ref{sec:multilabel}. Working with a much smaller number of ($K_\text{max}=3$) speakers opens up a new set of opportunities which have not yet been investigated by recent BoBW approaches. Inspired by~\cite{powerset_alibaba}, we propose to address \textit{speaker segmentation} as a frame-wise \textit{powerset} multi-class classification problem. Section~\ref{sec:powerset} provides a detailed description of the approach but one can understand the main idea by looking at Figure~\ref{fig:encoding}. The advantages of this approach are two-fold: it completely removes the need for the (sensitive) detection threshold hyperparameter, and it leads to significant performance improvement of the overall speaker diarization pipeline (as shown by a thorough evaluation on 9 benchmarking datasets in Sections~\ref{sec:experiments}~and~\ref{sec:results}). In particular, we claim state-of-the-art performance on AISHELL-4~\cite{AISHELL}, AliMeeting~\cite{AliMeeting}, AMI~\cite{AMI}, Ego4D~\cite{Ego4D}, MSDWild~\cite{MSDWild}, and REPERE~\cite{REPERE} benchmarks, make the approach available in the {\scriptsize \texttt{pyannote.audio}} library, and release pretrained models publicly.

\section{From multi-label classification...}
\label{sec:multilabel}

As depicted in the upper part of Figure~\ref{fig:encoding}, the reference segmentation of an audio chunk $\textbf{X}$ can be encoded into a sequence of $K_\text{max}$-dimensional binary frames $\mathbf{y} = \{ \mathbf{y_1}, \ldots, \mathbf{y_T}\}$ where $\mathbf{y_t} \in \{0, 1\}^{K_\text{max}}$ and $y_t^k= 1$ if speaker $k$ is active at frame $t$ and $y_t^k= 0$ otherwise. We may arbitrarily sort speakers by chronological order of their first activity but any permutation of the $K_\text{max}$ dimensions is a valid representation of the reference segmentation. Therefore, the binary cross entropy loss function $\mathcal{L}_\text{BCE}$ (classically used for such multi-label classification problems) has to be turned into a permutation-invariant loss function~$\mathcal{L}$ by running over all possible permutations $\text{perm}(\mathbf{y})$ of $\mathbf{y}$ over its $K_\text{max}$ dimensions:
\begin{equation}
\label{eq:pit}
\mathcal{L}\left(\mathbf{y}, \mathbf{\hat{y}}\right) = \min_{\text{perm}(\mathbf{y})} \mathcal{L}_\text{BCE} \left(\text{perm}(\mathbf{y}), \mathbf{\hat{y}}\right)
\end{equation}
with $\mathbf{\hat{y}} = f(\mathbf{X})$ where $f$ is the segmentation model whose architecture is described below. In practice, for efficiency, we first compute the $K_\text{max} \times K_\text{max}$ binary cross entropy losses between all pairs of $\mathbf{y}$ and $\mathbf{\hat{y}}$ dimensions, and rely on the Hungarian algorithm to find the permutation that minimizes the overall binary cross entropy loss. 

For the segmentation model $f$, we use the exact same architecture as the one introduced by~\cite{Bredin2021}. It consists in a \textit{SincNet} convolutional block~\cite{Ravanelli2018} (that performs frame-wise feature extraction), four bi-directional LSTMs (that contains most of the learnable weights), two fully-connected layers, a classification layer with $K_\text{max}$ outputs, and a final \textit{sigmoid} activation function to squash the frame-wise speaker activities $\hat{\mathbf{y}}$ between 0 and 1.

At test time, a subsequent binarization step is needed to decide whether each speaker $s$ is active in each frame $t$. This is achieved by comparing $\hat{\mathbf{y}}_{st}$ to a detection threshold $\theta \in \left[0, 1\right]$ whose value needs to be tuned carefully.

\section{... to powerset multi-class classification}
\label{sec:powerset}

While the standard multi-label loss relies on $K_\text{max}=3$ classes (one for each speaker $s_1$, $s_2$, $s_3$) that may be active simultaneously to encode overlapping speech, the \textit{powerset} multi-class encoding considers a grand total of $K_\text{powerset}=7$ mutually exclusive classes:
\begin{itemize}
    \item $\emptyset$ for non-speech frames;
    \item $\{s_1\}$, $\{s_2\}$, $\{s_3\}$ for frames with one active speaker;
    \item $\{s_1, s_2\}$, $\{s_1, s_3\}$, $\{s_2, s_3\}$ for two overlapping speakers\footnote{We do not consider the case of three or more overlapping speakers because the number of frames for which this happens in our benchmarking datasets is marginal: $1.6\%$ in the compound dataset, $0.73\%$ in DIHARD III.}.
\end{itemize}

Switching to \textit{powerset} encoding entails no radical change to the rest of the approach. It only amounts to changing the output size of the classification layer from $K_\text{max}=3$ to $K_\text{powerset}=7$, the final activation function from \textit{sigmoid} to \textit{softmax}, and the training loss function from binary to regular cross-entropy. Furthermore, a nice by-product is that the critical detection threshold~$\theta$ is removed in favor of a simple \textit{argmax} at test time.

Yet, \textit{powerset} permutation space is slightly more complex than in the multi-label setting. For example, swapping speakers $s_1$ and $s_2$ in the multi-label case obviously implies swapping classes $\{s_1\}$ and $\{s_2\}$, but also overlapping speakers classes $\{s_1,s_3\}$ and $\{s_2,s_3\}$. Therefore, to find the optimal permutation in the \textit{powerset} space, we use the following process:

\begin{itemize}
\item convert target ($\textbf{y}$) from \textit{powerset} to multi-label encoding;
\item convert binarized prediction ($\argmax_k \hat{\textbf{y}}$) from \textit{powerset} to multi-label encoding; 
\item find the optimal permutation as described in Section~\ref{sec:multilabel};
\item permutate the target accordingly in the multi-label space, and convert it back to \textit{powerset} encoding;
\item compute the cross-entropy loss in the \textit{powerset} space.
\end{itemize}

While~\cite{powerset_alibaba} introduced the use of \textit{powerset} encoding for speaker diarization, it has two main limitations. 
First, \textit{powerset} encoding does not play well with a variable number of speakers: adding an extra speaker to a pool of $K$ speakers results in $K + 1$ new \textit{powerset} classes (one for the actual speaker, and one for each possible overlapping speakers classes). Therefore, \cite{powerset_alibaba} had to resort to use a fixed and conservatively high number of speakers ($K_\text{max}=16$). Second, this leads to a number of \textit{powerset} classes which is prohibitively high. For instance, using $K_\text{max} = 16$ results in $K_\text{powerset} = \tbinom{0}{K} + \tbinom{1}{K} + \tbinom{2}{K} = 137$~classes, most of which are almost never seen during training. For reasons already explained in the introduction, our approach overcomes both issues: we can stick to a small upper bound on the number of speakers ($K_\text{max}=3$) thanks to our processing of short audio chunks, and the unsupervised clustering step takes care of estimating the number of speakers specific to each conversation later in the pipeline.

\section{Experiments}
\label{sec:experiments}

To ensure reproducibility, the research described in this paper builds on top of the speaker diarization pipeline from version {\scriptsize \texttt{2.1.1}} of {\scriptsize \texttt{pyannote.audio}} open source toolkit that follows the three-steps principle of BoBW approaches. More details on the pipeline internals are available in~\cite{pyannote2} but we describe the gist of it in the following section.

\subsection{Baseline}
\label{ssec:baseline}

The first step consists in applying the pretrained end-to-end neural speaker segmentation model {\scriptsize \texttt{pyannote/segmentation}} introduced in~\cite{Bredin2021} using a sliding window of 5s with a step of 500ms. A binarization step is further applied using the detection threshold $\theta \in \left[0, 1\right]$ which constitutes the first hyperparameter of the approach. 
The second step consists in extracting one speaker embedding per active speaker in each 5s~window. More precisely, speaker embeddings are only extracted from audio samples with exactly one active speaker: overlapping speech regions inferred automatically from the first step are discarded before computing the embeddings. All experiments reported in this paper rely on the pretrained ECAPA-TDNN model~\cite{ecapa} from {\scriptsize \texttt{SpeechBrain}}~\cite{speechbrain} available at {\scriptsize \texttt{hf.co/speechbrain/spkrec-ecapa-voxceleb}}.
The third and final step consists in applying agglomerative clustering on the aforementioned embeddings, using a second hyperparameter~$\delta$ as the maximum allowed distance between centroids of two clusters for them to be merged. Once each local (\textit{i.e.} from first step) active speaker is assigned to a global cluster, the global speaker diarization output can be constructed.

\subsection{Datasets}
\label{ssec:datasets}

\begin{figure}[t]
  \centering
  \includegraphics[width=\linewidth]{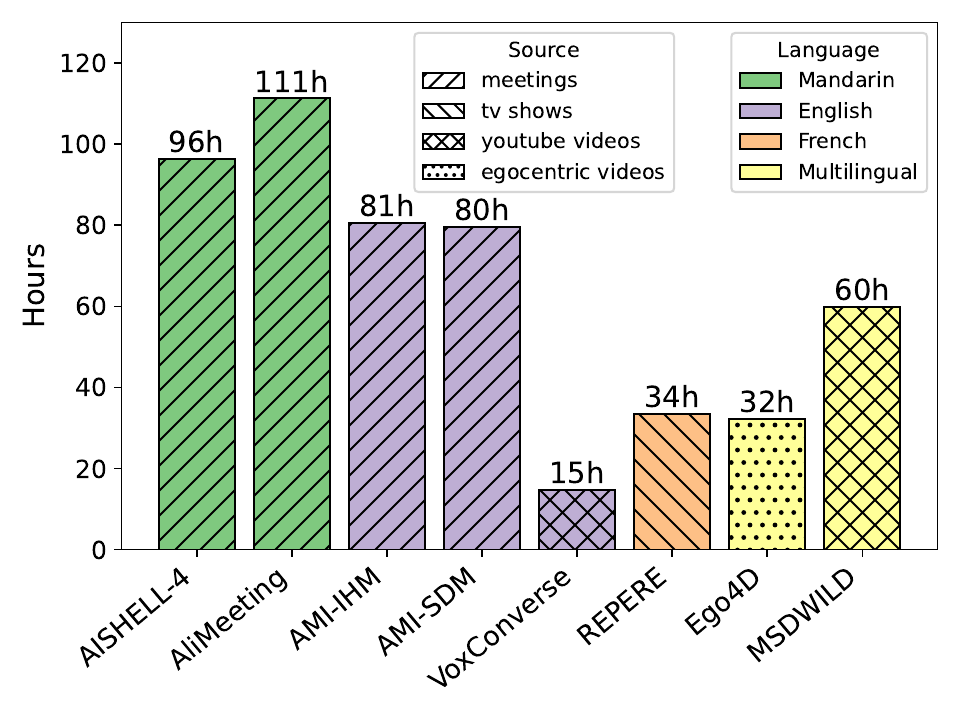}
  \caption{Compound multi-domain training set.}
  \label{fig:datasets}
\end{figure}

We report performance on the official test sets of 9 different datasets: AISHELL-4 \cite{AISHELL}, AliMeeting \cite{AliMeeting}, AMI \cite{AMI} (with two variants), DIHARD III~\cite{DIHARD}, Ego4D\footnote{At the time of writing this paper, test labels of Ego4D were not available. Hence, Ego4D results were obtained on its \textit{development} subset and should therefore be taken with a grain of salt.}~\cite{Ego4D}, MSDWild \cite{MSDWild}, REPERE \cite{REPERE}, and VoxConverse \cite{VoxConverse}. AMI variants are \textit{headset mix} (or \textit{IHM}, merged audio of the participants' headset microphones) and \textit{array 1 channel 1} (or \textit{SDM}, first channel from the first far-field microphone array).

For datasets distributed with an official training/development split, we use the training subset as is. For those only providing a development set, we split it into two parts which we call training and development subsets in the rest of the paper. Models were trained on a compound multi-domain training set made of the concatenation of the training sets of all aforementioned datasets, but DIHARD. DIHARD was kept aside to measure how robust our models are to domain mismatch: it is composed of 11 sub-domains, all tested individually in this paper. \autoref{fig:datasets} showcases the amount of data from each dataset in the compound training set as well as their source and language. We do expect that including DIHARD in the compound training set would significantly improve the results found in this paper.

While \autoref{fig:datasets} highlights the relative imbalance between each dataset in the compound training set, we do make sure that the compound development set is balanced, in two variants:
\begin{itemize}
    \item \smalltexttt{dev\_duration} is built by picking random 5s chunks from the \textit{development} subset of each dataset, for a grand total of 8~hours (one hour per dataset);
    \item \smalltexttt{dev\_files} is built by randomly picking 25 files from the \textit{development} subset of each dataset\footnote{Except for AISHELL-4, AliMeeting, and AMI which have only 20, 8, and 18 files in their respective development subsets.}, for a grand total of 164~files.
\end{itemize}

\subsection{Experimental protocol}
\label{ssec:protocol}

Speaker segmentation models are trained on the compound training set for at most one hundred hours, using Adam optimizer, an initial learning rate of $10^{-3}$, and a scheduler that divides the learning rate by $2$ after 30 epochs with no improvement. The validation metric used by the scheduler and for model selection is the local (\textit{i.e.} computed on 5s chunks) diarization error rate on the \smalltexttt{dev\_duration} development set.

hyperparameters tuning is performed with {\scriptsize \texttt{pyannote.pipeline}} in order to minimize the average diarization error rate on the \smalltexttt{dev\_files} compound development subset. More specifically, in the case of an internal multi-label speaker segmentation model, two hyperparameters need to be optimized: the detection threshold~$\theta$ and the clustering threshold~$\delta$. In the case of \textit{powerset} segmentation models, the detection threshold $\theta$ is not needed (because it is replaced by a parameter-less \textit{argmax}): only the clustering threshold~$\delta$ needs to be optimized.


\textit{Domain adaptation} experiments reported in Section~\ref{sec:results} are done by fine-tuning speaker segmentation models further on one specific domain. In those cases, for each dataset (or domain in case of DIHARD), model fine-tuning relies on the corresponding training set, model selection relies on the whole development set, and so does the tuning of pipeline hyperparameters.



\section{Results and discussions}
\label{sec:results}

\begin{table*}[!h]
  \caption{Performance on 8 different datasets. We report diarization error rates on the official test sets (with the exception of \textit{Ego4D} validation set because test labels are not available). \textbf{Best result for each dataset} are reported in bold (\textbf{as well as those less than 5\% worse relatively}). No forgiveness collar is used for evaluation, except for numbers in italics with a grey background \colorbox{gray!20}{computed with a 250ms forgiveness collar} (to allow comparison with some results reported in the literature). \faTrophy ~ indicates the winning (ensemble) submission to the VoxSRC or DIHARD III challenges.}
  \label{tab:results}
  \centering
  \begin{tabular}{l D{.}{.}{2.1} D{.}{.}{2.1} D{.}{.}{2.1} D{.}{.}{2.1} D{.}{.}{2.1} D{.}{.}{2.1} D{.}{.}{2.1} D{.}{.}{2.1}  D{.}{.}{2.1} D{.}{.}{2.1} D{.}{.}{2.1}}
\toprule & \multicolumn{1}{c}{\rot{\parbox{1.8cm}{\centering AISHELL-4\\\textit{channel 1}}}} & \multicolumn{1}{c}{\rot{\parbox{1.8cm}{\centering AliMeeting\\\textit{channel 1}}}} & \multicolumn{1}{c}{\rot{\parbox{1.8cm}{\centering AMI\\\textit{headset mix}}}} & \multicolumn{1}{c}{\rot{\parbox{1.8cm}{\centering AMI \textit{array 1}\\ \textit{channel 1}}}} & \multicolumn{1}{c}{\rot{\parbox{1.8cm}{\centering Ego4D \textit{v1}\\ \textit{validation}}}} &  \multicolumn{1}{c}{\rot{\parbox{1.8cm}{\centering MSDWild}}} & \multicolumn{1}{c}{\rot{\parbox{1.8cm}{\centering REPERE\\ \textit{phase 2}}}}  & \multicolumn{1}{c}{\rot{\parbox{1.8cm}{\centering VoxConverse\\ \textit{v0.3}}}} & \multicolumn{1}{c}{\rot{\parbox{1.8cm}{\centering \textit{in domain} \\ average}}} & \multicolumn{1}{c}{\rot{\parbox{1.8cm}{\centering DIHARD III}}}   \\
 & \multicolumn{1}{c}{\cite{AISHELL}} & \multicolumn{1}{c}{\cite{AliMeeting}} & \multicolumn{1}{c}{\cite{AMI}} & \multicolumn{1}{c}{\cite{AMI}} & \multicolumn{1}{c}{\cite{Ego4D}} & \multicolumn{1}{c}{\cite{MSDWild}} & \multicolumn{1}{c}{\cite{REPERE}} & \multicolumn{1}{c}{\cite{VoxConverse}} & & \multicolumn{1}{c}{\cite{DIHARD}} \\ 
\midrule
Pretrained {\scriptsize \texttt{pyannote/segmentation}} baseline~\cite{Bredin2021} & 14.7 & 26.1 & 19.0 & 29.1  & 63.2  & 34.8  & \textbf{8}.\textbf{5} & 11.9 & 25.9 & 23.7 \\
\midrule
Multi-label compound training & 17.1 & 27.2 & 22.3 & 24.7  & 58.7  & 33.5  & 9.0  & 12.1 &  25.6 & 33.8   \\
\variant with domain adaptation & 14.0 & 25.7 & 19.8 & 24.5 & 52.4 & \textbf{27}.\textbf{1}  & 9.1 & 11.3 & 23.0 & 22.4 \\
\cellcolor{gray!20} \subvariant \textit{with 250ms forgiveness collar} & \cellcolor{gray!20} \textit{8}.\textit{3} & \cellcolor{gray!20} \textit{15}.\textit{7} & \cellcolor{gray!20} \textit{12}.\textit{2} & \cellcolor{gray!20} \textit{16}.\textit{3} & \cellcolor{gray!20} \textit{43}.\textit{0} & \cellcolor{gray!20} \textbf{\textit{16}}.\textbf{\textit{0}} & \cellcolor{gray!20} \textit{6}.\textit{9} & \cellcolor{gray!20} \textit{6}.\textit{4} & \cellcolor{gray!20} \textit{15}.\textit{6} & \cellcolor{gray!20} \textit{12}.\textit{3} \\
\midrule
\textit{Powerset} compound training & 16.9 & \textbf{23}.\textbf{3} & 19.7 & \textbf{22}.\textbf{0}  & 57.2  & 29.2  & \textbf{8}.\textbf{4} & 11.6  & 23.5 & 29.9 \\
\variant with domain adaptation & \textbf{13}.\textbf{2} & 24.5 & \textbf{18}.\textbf{0} & \textbf{22}.\textbf{9}  & \textbf{48}.\textbf{2}  & 28.5  & \textbf{8}.\textbf{2}  & 10.4 & \textbf{21}.\textbf{7} & 21.3 \\
\cellcolor{gray!20} \subvariant \textit{with 250ms forgiveness collar} & \cellcolor{gray!20} \textbf{\textit{7}}.\textbf{\textit{6}} & \cellcolor{gray!20} \textit{15}.\textit{8} & \cellcolor{gray!20} \textbf{\textit{11}}.\textbf{\textit{2}} & \cellcolor{gray!20} \textbf{\textit{15}}.\textbf{\textit{3}} & \cellcolor{gray!20} \textbf{\textit{38}}.\textbf{\textit{9}}  & \cellcolor{gray!20} \textit{18}.\textit{6} & \cellcolor{gray!20} \textbf{\textit{6}}.\textbf{\textit{3}}  & \cellcolor{gray!20} \textit{5}.\textit{8} & \cellcolor{gray!20} \textbf{\textit{14}}.\textbf{\textit{9}} & \cellcolor{gray!20} \textit{12}.\textit{1}  \\
\midrule
\multicolumn{1}{r}{State of the art as of Feb. 2023} & 16.1 & \textbf{23}.\textbf{5} & 19.0 & 23.7 & 67.2 & \cellcolor{gray!20} \textit{22}.\textit{0} & 12.6 & \cellcolor{gray!20} \textbf{\textit{5}}.\textbf{\textit{1}} & & \textbf{16}.\textbf{8} \\
\multicolumn{1}{r}{[Source]} & \multicolumn{1}{c}{\cite{GPUGSS}} & \multicolumn{1}{c}{\cite{GPUGSS}} & \multicolumn{1}{c}{\cite{vbx}} & \multicolumn{1}{c}{\cite{GPUGSS}} & \multicolumn{1}{c}{\cite{Ego4D}} & \multicolumn{1}{c}{\cellcolor{gray!20} \cite{MSDWild}} & \multicolumn{1}{c}{\cite{Bredin2020}} & \multicolumn{1}{c}{{\cellcolor{gray!20} \faTrophy}} & & \multicolumn{1}{c}{\faTrophy}  \\
\bottomrule
  \end{tabular}
\end{table*}

\begin{figure*}[!h]
  \centering
  \includegraphics[width=\linewidth]{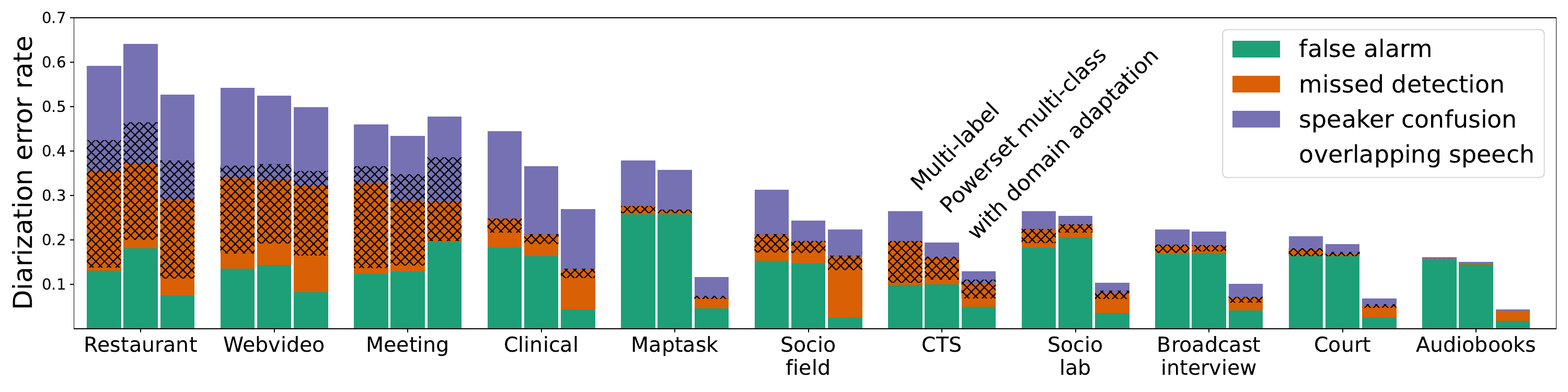}
  \caption{Decomposition of diarization error rate on DIHARD~III in terms of its false alarm, missed detection, and speaker confusion components. Errors happening in conjunction with overlapping speakers are hatched.}
  \label{fig:components_dihard}
\end{figure*}

While the first line of \autoref{tab:results} reports the performance of the speaker diarization pipeline based on the pretrained multi-label model available at {\scriptsize \url{hf.co/pyannote/segmentation}}, the second line is our (successful) attempt at retraining it from scratch with the proposed compound training set.

\textit{Powerset} compound training consistently gets better results than its multi-label counterpart -- with an average relative improvement of 8\% for  \textit{in domain} datasets (from 25.6\% to 23.5\% diarization error rate). Further adapting the \textit{powerset} model to each dataset gives an additional 8\% relative improvement, reaching state of the art performance for most of them.

Both approaches are tested for robustness to domain mismatch using the DIHARD~III dataset, which is composed of 11~domains that were never seen during (compound) training. \textit{Powerset} training shows a $11\%$ relative improvement over multi-label (from 33.8\% to 29.9\%), even larger than for the \textit{in domain} case.  This suggests that \textit{powerset} training is more robust to domain mismatch than its multi-label variant -- a behavior that could be explained by the removal of the otherwise sensitive detection threshold~$\theta$.


\autoref{fig:components_dihard} provides a detailed analysis of the errors committed by the various approaches. We find that most of the improvement comes from a consistent reduction in missed overlapping speech. This also holds for \textit{in-domain} data (with missed detection rate reduced from 13.1\% to 9.9\% on average), supporting the idea that \textit{powerset} explicit modeling of overlapping speakers classes has a huge impact on the overall performance.

Fine-tuning the \textit{powerset} segmentation model on each DIHARD domain (\textit{`with domain adaptation'} bars in \autoref{fig:components_dihard}) tends to converge towards an overall better compromise between false alarm and missed detection rates, with little to no impact on speaker confusion. This suggests that this additional domain adaptation step also plays the role of the detection threshold $\theta$ (which is a separate hyperparameter in the multi-label case); thus making the whole approach even more robust.

\section{Conclusion}

In this paper, we study the impact of framing the speaker diarization task as a \textit{powerset} multi-class classification problem. We extensively test this approach on 8 \textit{in domain} datasets, as well as the 11 domains of DIHARD used as \textit{out-of-domain} data. Compared to the classic multi-label approach, using a \textit{powerset} multi-class encoding results in significant diarization error rate improvement (mostly due to better predictions on overlapping speech) and better robustness to domain mismatch.  We obtain state of the art performance on AISHELL-4, AliMeeting, AMI, Ego4D, MSDWild, and REPERE. In the spirit of reproducible research, the \textit{powerset} multi-class segmentation code is available in the open-source {\scriptsize \texttt{pyannote.audio}} library. The models trained on the compound training dataset, as well as their precomputed output on each separate dataset are available at {\scriptsize\url{github.com/FrenchKrab/IS2023-powerset-diarization}}.
\ifinterspeechfinal

\section{Acknowledgements}
This work was granted access to the HPC resources of IDRIS under the allocations AD011013477 and AD011012177R2 made by GENCI.

\else

\fi

\bibliographystyle{IEEEtran}
\bibliography{reference}

\end{document}